\documentclass[runningheads]{llncs}
\usepackage[T1]{fontenc}
\usepackage{hyperref}
\usepackage{graphicx,verbatim}
\usepackage{amsmath}
\usepackage{bm}
\usepackage{amssymb}

\usepackage{booktabs}
\usepackage[table]{xcolor}
\usepackage{amssymb}
\usepackage{multirow}
\usepackage{graphicx}
\begin{document}

\title{SpecF2M: A Spectral-Aware Multi-task Network Estimating Axial Length and Refractive Error from Pediatric Fundus Photographs}
\titlerunning{SpecF2M}
\authorrunning{M. He et al.}

\author{\author{Mengxian He\inst{1}$^{\star}$\orcidID{0009-0005-7585-374X}  \and Xinyue Liu\inst{2}\orcidID{0009-0003-6294-9932}\thanks{Contributed equally to this work.} \and Yunyun Sun\inst{3} \and Hao Wei\inst{1}\orcidID{0000-0002-5719-8826} \and Minqing Zhang\inst{1}\orcidID{0000-0002-7214-0569} \and Shunyi Zhang\inst{3} \and Lichun Wang\inst{4} \and Wu Yuan\inst{1}\orcidID{0000-0001-9405-519X}\thanks{Corresponding author}}}


\institute{The Chinese University of Hong Kong, Sha Tin, New Territories, Hong Kong \\
    \email{wyuan@cuhk.edu.hk} \and 
    Guangdong Laboratory of Machine Perception and Intelligent Computing, Shenzhen MSU-BIT University, China \and
    Capital Medical University, China \and
    Zhengzhou Second Hospital, China
}
  
\maketitle              

 \begin{abstract}

Spherical Equivalent Refraction (SER) and Axial Length (AL) are core indicators for pediatric myopia screening, yet their measurements require dedicated biometry and cycloplegic refraction. Fundus photography offers an accessible imaging modality, as myopia-related posterior-pole changes are visible in 45$^\circ$ fundus images. However, these cues are often low-contrast, spatially diffuse, and multi-scale. Moreover, AL, Sphere (SPH), and Cylinder (CYL) share partially overlapping but non-identical anatomical correlates. We propose SpecF2M, a spectral-aware multi-task network for estimating AL and SER components from pediatric fundus photographs. SpecF2M integrates a deterministic anatomy-guided enhancement module, a hybrid spatial--spectral backbone combining MixCNN and Hybrid Spectral Learning (HSL) blocks, and an expert-routing head for component-level estimation of AL, SPH, and CYL. On a pediatric cohort of 4,359 eligible child visits and 6,966 fundus images, SpecF2M outperforms controlled CNN/ViT baselines for AL and SPH estimation, achieving MAEs of 0.5347 mm and 0.7062 D, respectively. Component-level analysis further reveals asymmetric task coupling, where CYL exhibits weaker association with fundus-derived myopic patterns than AL/SPH. These results support fundus-based, screening-oriented estimation of pediatric myopia indicators, while external validation remains necessary before deployment.


\end{abstract}

\section{Introduction}

Myopia is a global public health burden, with a high and increasing prevalence among children \cite{morgan2018epidemics}. Early-onset myopia tends to progress rapidly, substantially increasing the risk of vision-threatening complications \cite{hu2020association}. Among school-aged children, axial myopia is the most prevalent subtype, characterized by excessive elongation of the ocular axis, which shifts the retinal plane posterior to the focal plane and results in a progressive myopic refractive error \cite{jonas2023imi}. 

Accordingly, in clinical practice, Spherical Equivalent Refraction (SER) and Axial Length (AL) are critical quantitative indicators to assess myopia severity and progression risk. SER is computed as $SER = SPH + 0.5 \times CYL$ \cite{enaholo2023spherical}, where SPH (Sphere) denotes the spherical refractive component, and CYL (Cylinder) denotes the astigmatic component. AL characterizes the structural growth of the eye along the anterior–posterior axis. Consequently, myopia progression in children is largely driven by AL elongation \cite{flitcroft2019imi}, whereas CYL remains relatively stable, suggesting that SER changes are mainly attributable to the SPH \cite{harvey2015longitudinal}.

Therefore, measurements of AL and SER are crucial for screening pediatric myopia and estimating its risk of progression. However, obtaining SER requires cycloplegic refraction, which is time-consuming and often uncomfortable for children, whereas AL measurement depends on dedicated devices (e.g., Lenstar900). These required workflows, equipment, manpower, and high compliance limit accessibility and scalability, hindering large-scale screening and management of pediatric myopia.

\begin{figure}
\centering
\includegraphics[width=1.0\linewidth]{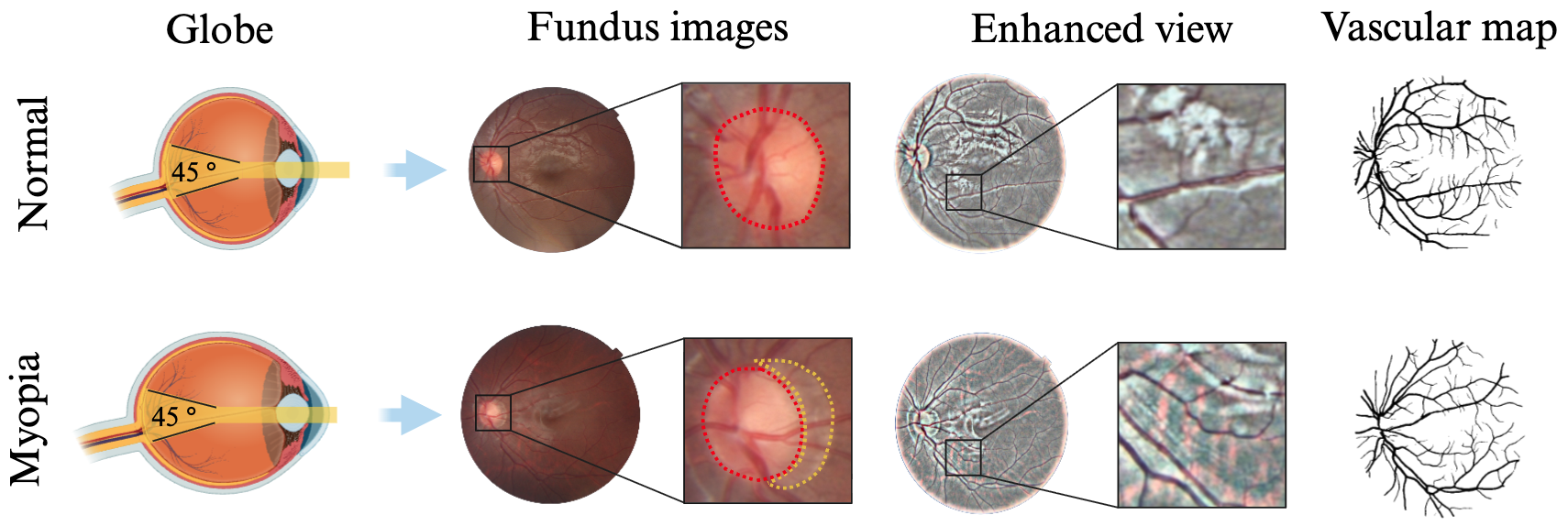}
\caption{Examples of a normal eye (AL 20.73 mm, SER $+1.375\,\mathrm{D}$) and a myopic eye (AL 26.11 mm, SER $-6.375\,\mathrm{D}$), showing anatomical regions related to OD, PPA, FT, and FV. } \label{fig_1}
\end{figure}

In contrast, 45° posterior-pole fundus photography (45° fundus images) is widely used in screening settings \cite{williams2004single}, offering a lower-cost, objective, and repeatable examination. Prior studies have shown that pediatric myopia progression is accompanied by characteristic posterior-pole changes that are visible on 45° fundus images \cite{yii2024retinal}. As illustrated in Fig.~\ref{fig_1}, elongation of the ocular axis mechanically stretches the posterior globe, which gives rise to a set of common manifestations including (1) Optic Disc (OD) deformation or tilt \cite{kim2012optic}, (2) the progression of Peripapillary Atrophy (PPA) around OD \cite{kim2012optic}, (3) Fundus Tessellation (FT) due to retina–choroid thinning with increased choroidal vessel visibility \cite{ohno2015international}, and (4) straighter, elongated Fundus Vessel (FV) with reduced tortuosity \cite{he2024deep}. Together, the above changes combine global shape deformation with localized edge/texture alterations. These multi-scale features align naturally with spectral representations, which can capture myopic fundus changes more explicitly. 

Recent studies have shown that SER and AL can be estimated from fundus images. Early work demonstrated DL-based refractive-error estimation and found SPH to be more predictable than CYL~\cite{varadarajan2018deep}. Other studies estimated AL from fundus photographs or UWF images and showed that AL-related cues are spatially diffuse and scale-dependent~\cite{dong2021deep,jeong2020ocular,yang2024convolutional}. Joint AL/SER modeling has also been explored using statistical or deep-learning formulations~\cite{oh2023deep,li2024oucopula,zhong2025cecnn}. However, existing methods mainly rely on coarse joint prediction or conventional CNN features, leaving component-aware coupling, anatomy-guided enhancement, and spectral-aware representation insufficiently explored.

\textit{SpecF2M} addresses these gaps from three perspectives:

\begin{enumerate}
    \item \textbf{Expert-routing multi-task learning.} We formulate AL, SPH, and CYL estimation as a component-level multi-task problem and use expert routing to balance shared and task-specific representations.
    \item \textbf{Anatomy-guided enhancement strategy.} We introduce a training-free anatomy-guided enhancement strategy to highlight low-contrast myopic structures, including OD/PPA, tessellation, and vessel-related patterns.
    \item \textbf{Spectral-aware representation learning.} We design a spectral-aware backbone that combines local multi-resolution wavelet modeling with global Fourier-domain propagation.
\end{enumerate}

To our knowledge, \textit{SpecF2M} is the first framework to jointly estimate AL, SPH, and CYL from 45$^\circ$ pediatric fundus photographs, providing both screening-oriented prediction and insight into asymmetric AL/SPH/CYL coupling.

\section{Problem formulation}

Let $\mathcal{D}=\{(x_i,y_i^{AL},y_i^{SPH},y_i^{CYL})\}_{i=1}^{N}$ denote a pediatric fundus dataset, where $x_i$ is a fundus image and $y_i^{AL}$, $y_i^{SPH}$, and $y_i^{CYL}$ are the paired Axial Length (AL), Sphere (SPH), and Cylinder (CYL), respectively. We learn a multi-output regression model:
\[
(\hat{y}_i^{AL},\hat{y}_i^{SPH},\hat{y}_i^{CYL})=g_{\phi,\psi,\gamma}(h_{\theta}(x_i)),
\]
where $h_{\theta}(\cdot)$ is a shared backbone and $g_{\phi,\psi,\gamma}(\cdot)$ denotes task-specific heads. The predicted SER is computed as 
$\mathrm{SER}_{\mathrm{pred}}=\hat{y}_{\mathrm{SPH}}+0.5\times\hat{y}_{\mathrm{CYL}}$. We explicitly predict SPH and CYL rather than using SER as an auxiliary loss because SER is linearly determined by SPH/CYL and may mask component-specific errors when their residuals have opposite signs.

\section{Methods}
\begin{figure}[t]
    \centering
    \includegraphics[width=0.9\linewidth]{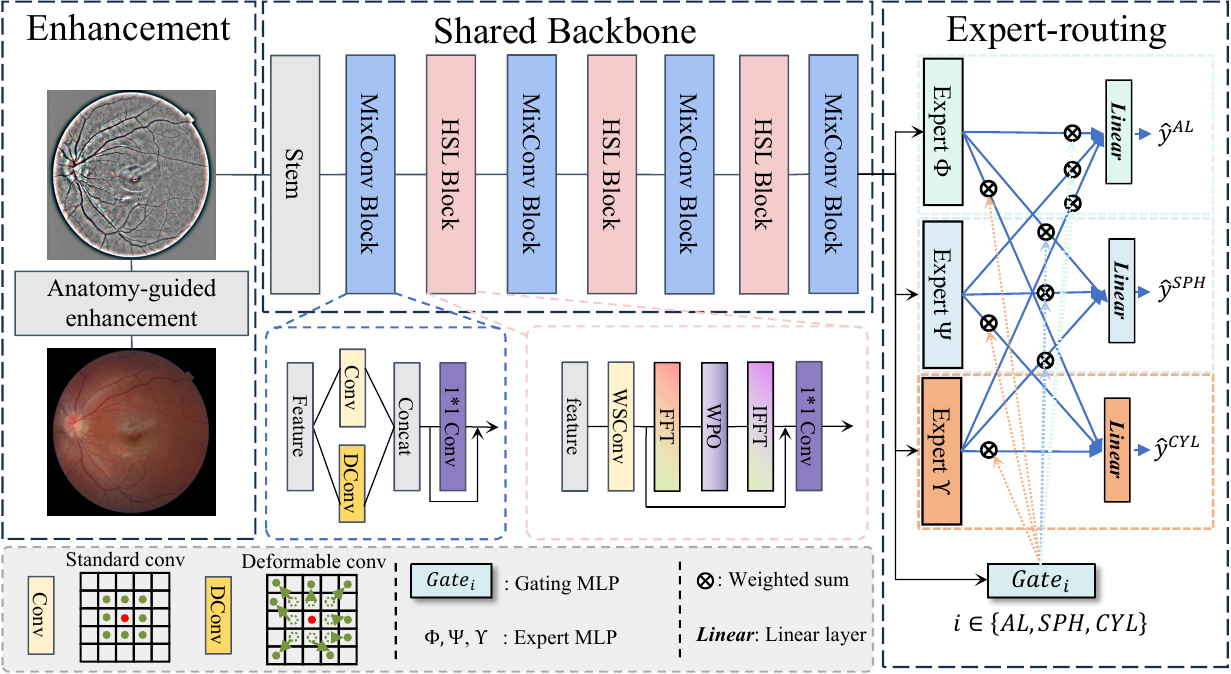}
    \caption{The overview of proposed \textit{SpecF2M}.}
    \label{fig_method}
\end{figure}

As shown in Fig.~2, SpecF2M contains three modules: anatomy-guided enhancement, a shared MixCNN--HSL backbone, and expert-routing regression heads. The enhanced image is first transformed into spatial--spectral features, which are then routed to expert-specific embeddings for AL, SPH, and CYL estimation.

\subsection{Anatomy-guided enhancement strategy}

As shown on the left of Fig. 2, we propose an anatomy-guided enhancement strategy for improving the discriminability of low-contrast, spatially diffuse myopic anatomical structures.
Given a color fundus image $I\in\mathbb{R}^{H\times W\times 3}$, we define a deterministic enhancement operator $\mathcal{A}(\cdot)$:
\begin{equation}
\tilde I=\mathcal{A}(I)=\mathcal{F}_{\text{roi}}\Big(\mathcal{F}_{\text{v-sup}}\big(\mathcal{F}_{\text{hf}}(\mathcal{F}_{\text{lum}}(I))\big)\Big).
\label{eq:aug_overall}
\end{equation}
Each component of $\mathcal{A}(\cdot)$ uses fixed hyperparameters and introduces no additional learnable parameters.

\textbf{(1) Luminance enhancement} ($\mathcal{F}_{\text{lum}}$).
We first resize and convert $I$ to LAB space $I^{lab}=\Phi_{\text{LAB}}(I)$, and apply CLAHE only on the luminance channel:
\begin{equation}
I_1=\mathcal{F}_{\text{lum}}(I)=\Phi_{\text{LAB}}^{-1}\big([\mathrm{CLAHE}(L),A,B]\big).
\label{eq:F_lum}
\end{equation}

\textbf{(2) High-frequency emphasis} ($\mathcal{F}_{\text{hf}}$).
To reduce low-frequency redundancy and highlight micro-structures, we apply an adaptive high-pass sharpening:
\begin{equation}
I_2=\mathcal{F}_{\text{hf}}(I_1)=I_1+\alpha\big(I_1-G_{\sigma}(I_1)\big),
\label{eq:F_hf}
\end{equation}
where $G_{\sigma}(\cdot)$ denotes Gaussian smoothing and $\alpha$ controls the sharpening strength.

\textbf{(3) Vessel-prior soft suppression} ($\mathcal{F}_{\text{v-sup}}$).
Since contrast enhancement can amplify vascular linear patterns, we derive a continuous soft vessel mask $M_v\in(0,1)^{H\times W}$ from the green channel $G$ via multi-orientation black-hat filtering:$ V = \sum_{\theta \in \Theta} \mathrm{BH}(G; SE_\theta), \quad M_v = \sigma\left( \frac{V - \mu(V)}{s(V)} \right),$ and suppress vessel textures in a soft manner:
\begin{equation}
I_3=\mathcal{F}_{\text{v-sup}}(I_2)=I_2-\lambda\,(M_v\odot I_2),
\label{eq:F_vsup}
\end{equation}
which increases the relative saliency of diffuse tessellation and structural details.

\textbf{(4) OD/PPA ROI enhancement with soft fusion} ($\mathcal{F}_{\text{roi}}$).
We obtain an optic-disc ROI mask $M_{od}=\mathcal{R}(I_3)$ by deterministic morphology and geometric fitting, enhance the ROI by a lightweight operator $\mathcal{E}(\cdot)$ (contrast redistribution + mild sharpening), and softly fuse it back to avoid boundary artifacts:
\begin{equation}
I_{od}=\mathcal{E}(I_3\odot M_{od}),
\qquad
W=G_{\tau}(M_{od}),
\qquad
\tilde I=\mathcal{F}_{\text{roi}}(I_3)=(1-W)\odot I_3+W\odot I_{od},
\label{eq:F_roi}
\end{equation}
where $G_{\tau}(\cdot)$ is the ROI mask to form a continuous blending map $W\in[0,1]$.

\subsection{Spectral-aware representation learning}

Myopic fundus changes are inherently multi-scale and spectrally structured, motivating a spectral-aware representation. Inspired by \cite{finder2024wavelet,shu2026waveformer}, we construct a shared backbone by interleaving MixCNN blocks with HSL blocks. This design progressively injects spectral information while preserving spatial information.

In the MixCNN block, we adopt a parallel design that combines the standard convolution and deformable convolution kernels to learn complementary spatial features. Given an input feature map $\mathbf{X} \in \mathbb{R}^{B \times C \times H \times W}$, the standard branch produces $\mathbf{F}_{\text{std}} = \mathrm{Conv}(\mathbf{X}),$ which focuses on learning stable local textures and regular structural patterns. 
In parallel, the deformable branch applies adaptive sampling to account for geometric deformations: $\mathbf{F}_{\text{def}} = \mathrm{DConv}(\mathbf{X}),
$ where the sampling offsets are learned to better align with irregular anatomical structures. The outputs are concatenated along the channel dimension and fused by a $1 \times 1$ convolution $\mathbf{F}_{\text{fuse}} = \mathrm{Conv}_{1 \times 1}\big([\mathbf{F}_{\text{std}}, \mathbf{F}_{\text{def}}]\big),$ providing a robust spatial representation for subsequent HSL blocks.

The HSL blocks further extend the representation into the spectral domain to learn multi-scale wavelet components and global frequency interactions. We construct a hybrid spectral learning block by cascading a modified \cite{finder2024wavelet} Wavelet-Selective CNN layer (WSConv) and a Fourier-domain global Wave Propagation Operator (WPO) \cite{shu2026waveformer}.
We first apply an $\ell$-level 2D wavelet transform (WT) per channel to obtain multi-resolution subbands.
Let $\mathcal{W}(\cdot)$ and $\mathcal{W}^{-1}(\cdot)$ denote WT and inverse WT (IWT), respectively.
For level $i=1,\dots,\ell$, we decompose the current low-frequency band as
\begin{equation}
(\mathbf{X}^{(i)}_{LL},\, \mathbf{X}^{(i)}_{H}) = \mathcal{W}\!\left(\mathbf{X}^{(i-1)}_{LL}\right), 
\quad \mathbf{X}^{(0)}_{LL}=\mathbf{X},
\end{equation}
where $\mathbf{X}^{(i)}_{H}$ stacks the high-frequency subbands $\{LH, HL, HH\}$.
We then perform subband-specific convolutions:
\begin{equation}
\mathbf{Y}^{(i)}_{LL}=\mathrm{Conv}^{(i)}_{LL}\!\left(\mathbf{X}^{(i)}_{LL}\right),\qquad
\mathbf{Y}^{(i)}_{H}=\mathrm{Conv}^{(i)}_{H}\!\left(\mathbf{X}^{(i)}_{H}\right).
\end{equation}
Given that enhancement attenuates LL content, we compute a lightweight gate $g$ from pooled LH responses via an MLP and sigmoid activation. We subsequently use the LH-guided gate to modulate the LL response and add it back as a residual. Following the linearity of IWT, WSConv aggregates outputs across levels via a recursive reconstruction:
\begin{equation}
 \mathbf{Z}^{(\ell+1)}=\mathbf{0},
\mathbf{Z}^{(i)}=\mathcal{W}^{-1}\!\left(\mathbf{Y}^{(i)}_{LL}+\mathbf{Z}^{(i+1)},\, \mathbf{Y}^{(i)}_{H}\right),
\end{equation}
 where $i=\ell,\dots,1$, and produces the feature: $\mathbf{U}_0 = \mathrm{WSConv}(\mathbf{X}).$

We next propagate $\mathbf{U}_0$ globally with WPO.
Let $\mathcal{F}(\cdot)$ and $\mathcal{F}^{-1}(\cdot)$ denote the 2D FFT/ IFFT applied per channel.
WPO evolves $\mathbf{U}_0$ over the internal time $t$ under the underdamped wave dynamics in the frequency domain:
\begin{equation}
\mathbf{U}_t
= \mathcal{F}^{-1}\!\left\{
e^{-\frac{\alpha}{2}t}\left[
\mathcal{F}(\mathbf{U}_0)\cos(\omega_d t)
+\frac{\sin(\omega_d t)}{\omega_d}\Big(\mathcal{F}(\mathbf{V}_0)+\frac{\alpha}{2}\mathcal{F}(\mathbf{U}_0)\Big)
\right]
\right\},
\end{equation}
where $\alpha$ is the damping coefficient, $v$ is the wave speed (encoded in $\omega_d$), and $\mathbf{V}_0$ is the initial velocity.
In practice, we take the real part after IFFT: $\tilde{\mathbf{U}} = \Re(\mathbf{U}_t)$, and set $\alpha=0.1$ and $v=1$ as the starting point, and set $t$ according to the depth of the HSL block. Finally, we fuse the globally propagated feature with the WSConv output via a skip connection and a $1\times1$ convolution:
$ \mathbf{Y} = \mathrm{Conv}_{1\times 1}\!\left(\tilde{\mathbf{U}} + \mathbf{U}_0\right).$ 

In this work, HSL is treated as an integrated spectral block that combines local multi-resolution subband modeling and global frequency-domain propagation; our empirical claim is the effectiveness of this coupled design rather than complete causal attribution of each individual spectral operator.

\subsection{Expert-routing multi-task learning}
We formulate AL together with SPH and CYL as a SER-component-level multi-task setup.

Given an input enhanced fundus image $\tilde I$, the preceding modules output a shared feature representation $\mathbf{Y}$.
On top of $\mathbf{Y}$, we adopt the experts routing scheme to balance shared representation learning and task-specific disentanglement for three regression tasks: $\mathcal{T}=\{\mathrm{AL},\mathrm{SPH},\mathrm{CYL}\}.$

We instantiate three experts $\{g_{\phi}, g_{\psi}, g_{\gamma}\} \in  g_{\phi,\psi,\gamma}$ that transform the shared feature into expert-specific embeddings: $\mathbf{e}_{k} = g_{k}(\mathbf{Y}), \quad k\in\{\phi,\psi,\gamma\}$.
where $\mathbf{e}_{k}\in\mathbb{R}^{d}$ denotes the expert feature. All experts take the same input $\mathbf{Y}$ but are parameterized independently, allowing diversity.

For each task $t\in\mathcal{T}$, a dedicated gate $G_{t}$ predicts routing weights over experts: $\boldsymbol{\alpha}_{t}=\mathrm{softmax}\!\left(G_{t}(\mathbf{Y})\right)\in\mathbb{R}^{3}, \quad \sum_{k}\alpha_{t,k}=1.$ The routed task feature is obtained via a convex combination of expert embeddings: $
\mathbf{z}_{t}=\sum_{k\in\{\phi,\psi,\gamma\}}\alpha_{t,k}\,\mathbf{e}_{k}.$ Each routed feature $\mathbf{z}_{t}$ is fed into a lightweight linear regressor to produce the estimation: $ \hat{y}_{t}= \mathbf{w}_{t}^{\top}\mathbf{z}_{t}+b_{t}, t\in\mathcal{T},$ yielding $(\hat{y}_i^{AL}, \hat{y}_i^{SPH}, \hat{y}_i^{CYL})$.

We optimize a weighted multi-task regression objective:
\begin{equation}
\mathcal{L}=\sum_{t\in\mathcal{T}}\lambda_{t}\,\ell\!\left(\hat{y}_{t},y_{t}\right),
\end{equation}
where $\ell(\cdot)$ is a regression loss and $\lambda_{t}$ balances task contributions.

\section{Experiments}
The data were collected from a pediatric myopia cohort across seven primary schools, including 4,359 eligible child visits. After excluding images with unclear optic disc, strong reflection, or artifacts, 6,966 fundus images with paired AL and cycloplegic refraction measurements were included. The mean age was 8.22$\pm$1.09 years; mean AL, SPH, and CYL were 23.88$\pm$1.07 mm, -0.19$\pm$1.85 D, and -0.83$\pm$0.95 D, respectively. Cycloplegia was induced with 1\% cyclopentolate hydrochloride eye drops before autorefraction. The study was approved by TRECKY2019-058. Images and clinical measurements were de-identified. Written informed consent was provided by parents or legal guardians, and verbal assent was obtained from children when applicable.

\subsection{Evaluation setup}
\noindent\textbf{Evaluation.} We split the data by subject ID to avoid inter-eye leakage, keeping both eyes of the same subject in the same split. Training, validation, and test sets followed an 80/10/10 ratio and were stratified by myopia level. Performance was evaluated using Mean Absolute Error (MAE) and $R^2$ for SPH, CYL, and AL.

\noindent\textbf{Baseline.} We compared a range of representative baseline methods covering the ViT and CNN families in both single-task and standard hard multi-task settings and reported the best performance.

\noindent\textbf{Implementation.} SpecF2M was trained on an NVIDIA RTX 4090D GPU for 150 epochs with a batch size of 32, a learning rate of $10^{-4}$, and a weight decay of $3\times10^{-5}$. We used MSE loss with task weights AL:SPH:CYL = 0.5:1:1 and geometric augmentations including random rotation, flipping, and mirroring. WSConv used a Daubechies-1 wavelet with a 2-level decomposition. Each expert MLP was a one-layer 512-d MLP with ReLU and dropout 0.2.

\begin{table}[htbp]
\centering
\scriptsize
\caption{Comparison with baselines and ablation study of \textit{SpecF2M}.}
\setlength{\tabcolsep}{4pt}
\renewcommand{\arraystretch}{1.05}

\resizebox{\linewidth}{!}{%
\begin{tabular}{p{2.5cm}| cc | cc | cc}
\toprule \rowcolor{white!5}
\multirow{2}{*}{Method} &
\multicolumn{2}{c|}{SPH} &
\multicolumn{2}{c|}{CYL} &
\multicolumn{2}{c}{AL} \\
& MAE & $R^2$ & MAE & $R^2$ & MAE & $R^2$ \\
\midrule

ResNet18 \cite{he2016deep}            & 0.8216 & 0.6445 & 0.5777 & 0.3995 & 0.8140 & 0.6029 \\
DenseNet121 \cite{huang2017densely}   &  0.7678 & 0.6607 & 0.5673 & 0.4030 & 0.5903 & 0.5469 \\
VGG16-BN \cite{simonyan2014very}      & 0.7715 & 0.6662 & 0.5527 & 0.4216 & 0.5646 & 0.5689 \\
ViT-S/16 \cite{touvron2021training}   & 0.9101 & 0.5132 & 0.6228 & 0.1222 & 0.7618 & 0.2498 \\
ViT-B/16 \cite{dosovitskiy2020image}  & 0.7710 & 0.6647 & \textbf{0.5178} & \textbf{0.4887} & 0.5787 & 0.5669 \\
\midrule
\textit{SpecF2M} (Our) & \textbf{0.7062} & \textbf{0.7251} & 0.5508 & 0.4632 & \textbf{0.5347} & \textbf{0.6038} \\
w/o Enhanced & 0.7547 & 0.6704 & 0.5433 & 0.4676 & 0.5703 & 0.5457 \\
 w/o HSL block & 0.8233 & 0.6341 & 0.5596 & 0.4505 & 0.6053 & 0.5134 \\
 w/o Expert-routing     & 0.8674 & 0.5601 & 0.5295 & 0.3634 & 0.6202 & 0.4958 \\

\bottomrule
\end{tabular}%
}
\label{tab:compare_ablation_merged}
\end{table}

\subsection{Main results}
Table~1 compares SpecF2M with controlled CNN/ViT baselines and ablations. SpecF2M achieves the best AL and SPH estimation performance, with MAE/$R^2$ of 0.5347/0.6038 for AL and 0.7062/0.7251 for SPH. 

\begin{figure}[h!]
    \centering
    \includegraphics[width=1.0\linewidth]{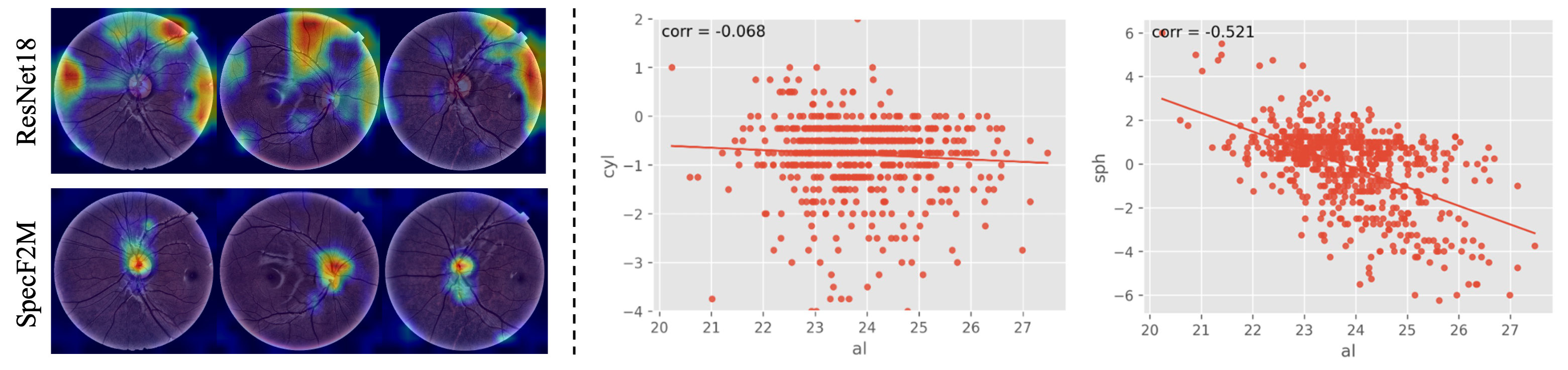}
    \caption{The Grad-CAM and task correlation analysis.}
    \label{fig_vis}
\end{figure}

Grad-CAM visualizations in Fig.~3 show that SpecF2M produces more compact responses around the OD and adjacent vascular structures than ResNet18. Ablations show that enhancement, HSL, and expert routing consistently benefit AL and SPH, whereas CYL changes are less stable. This is consistent with the correlation analysis in Fig.~3: AL is correlated with SPH (corr=$-0.521$) but nearly uncorrelated with CYL (corr=$-0.068$), suggesting weaker fundus-derived coupling for CYL.

\section{Discussion and Conclusion}

We propose SpecF2M for screening-oriented joint estimation of pediatric AL and refraction components from widely available 45$^\circ$ fundus photographs, highlighting its potential as an accessible image-based decision-support approach for pediatric myopia assessment. While \textit{SpecF2M} consistently outperforms controlled CNN/ViT baselines and ablations at AL and SPH estimation, the component-wise results further reveal an asymmetric coupling: AL and SPH benefit more reliably from multi-task learning, supporting stronger shared anatomical correlates and cross-task synergy, whereas CYL remains noisier and more variable, indicating higher stochasticity from the estimation based on fundus patterns.

\section*{Acknowledgements}
This work was supported in part by the Research Grants Council (RGC) of Hong Kong SAR (GRF14213125, GRF14201824, GRF14216222), and the Innovation and Technology Fund (ITF) of Hong Kong SAR (ITS/252/23).

\section*{Disclosure of Interests}
The authors have no competing interests to declare that are relevant to the content of this article.

\bibliographystyle{splncs04}
\bibliography{references}
\end{document}